\documentclass[preprint2]{aastex}

\shorttitle{A Time Delay for SDSS~J1029+2623}
\shortauthors{Fohlmeister et al. 2013}

\begin{document}

\title{A TWO-YEAR TIME DELAY FOR THE LENSED QUASAR SDSS J1029+2623}

\author{Janine Fohlmeister\altaffilmark{1}}
\affil{Astronomisches Rechen-Institut, Zentrum f\"ur Astronomie der Universit\"at Heidelberg, M\"onchhofstr. 12-14, D-
69120 Heidelberg, Germany}

\author{Christopher S. Kochanek\altaffilmark{2,}\altaffilmark{3}}
\affil{Department of Astronomy, The Ohio State University, Columbus, OH 43210, USA}
\affil{Center for Cosmology and Astroparticle Physics, The Ohio State University, 191 West Woodruff Avenue, Columbus, OH 43210, USA}

\author{Emilio E. Falco\altaffilmark{4}}
\affil{Harvard-Smithsonian Center for Astrophysics, Cambridge, MA02138, USA}

\author{Joachim Wambsganss\altaffilmark{1}}

\author{Masamune Oguri\altaffilmark{5}}
\affil{Kavli Institute for the Physics and Mathematics of the Universe, Todai Institutes for Advanced Study, University of Tokyo, Kashiwa, Chiba 277-8583, Japan}

\author{Xinyu Dai\altaffilmark{6}}
\affil{Department of Physics and Astronomy, University of Oklahoma, 440 W. Brooks Street, Norman, OK 73019, USA}

\begin{abstract}
 We present 279 epochs of optical monitoring data spanning 5.4 years
 from 2007 January to 2012 June for the largest image separation (22\farcs6) 
 gravitationally lensed quasar,
 SDSS J1029+2623.  We find that image A leads the images B and C by 
 $\Delta t_{AB} = (744 \pm 10)$~days (90\% confidence); the uncertainty includes both
 statistical uncertainties and systematic differences due to the
 choice of models.  With only a $\sim 1\%$ fractional error, the interpretation of the delay is limited 
 primarily by cosmic variance due to 
 fluctuations in the mean line-of-sight density. 
 We cannot separate the fainter image C from image B, but since image
 C trails image B by only 2--3 days in all models, the estimate
 of the time delay between image A and B is little affected by combining the fluxes
 of images B and C. There is weak evidence for a low level of
 microlensing, perhaps created by the small galaxy 
 responsible for the flux ratio anomaly in this system.  
 Interpreting the
 delay depends on better constraining the shape of the 
 gravitational potential using the lensed host galaxy, other
 lensed arcs and the structure of the X-ray emission. 
\end{abstract}

\keywords{galaxies: clusters: general --- gravitational lensing: strong ---
quasars: individual (SDSS J102913.94+262317.9)}

\section{Introduction}

SDSS~J1029+2623 \citep{Inada2006} is the largest image separation lensed quasar, with 
a maximum separation of 22\farcs6 that significantly exceeds that of the next 
largest system (14\farcs6 for SDSS~J1004+4112; \citealt{Inada2003}).  Although
it was first identified with only two images A and B, it actually consists of
three images of a $z_s=2.197$ quasar produced by a $z_l=0.58$ galaxy cluster
in a rare ``naked cusp'' configuration (\citealt{Oguri2008}). The
fainter image C lies close to image B (1\farcs8), which would usually mean that
B and C should be significantly brighter than A.  Instead, the optical
flux ratios of the images, A:B:C=0.95:1.00:0.24, show a large anomaly
that cannot be reproduced by ellipsoidal models centered near the
bright cluster galaxies (\citealt{Oguri2008}).  The quasar is radio
loud, and the flux ratio anomaly persists in the radio, albeit with
different flux ratios than in the optical (\citealt{Kratzer2011}). 
Recent \textit{Chandra} X-ray observations (\citealt{Ota2012}) find a cluster
mass consistent with lens models and that there is soft X-ray absorption in
the spectrum of image C consistent with explaining the optical color differences between
the images A, B and C as extinction.  After correction for this absorption/extinction, the
X-ray, optical and radio flux ratios are broadly consistent and the flux anomalies are due to a small galaxy near 
image C (\citealt{Oguri2012b}). 

As part of a program to better understand this system, including deep \textit{Hubble Space Telescope} 
(\citealt{Oguri2012b}), X-ray (\citealt{Ota2012}), radio (\citealt{Kratzer2011}), 
and weak-lensing (\citealt{Oguri2012a}) observations, we have 
been monitoring the lens in the optical since 2007 to measure the time delay.  Time delays
generally measure a combination of cosmological distances (the
Hubble constant to lowest order) and the surface density of the 
lens at the radius of the images (\citealt{kochanek2002}). Since
the mass distribution of the lens can be independently constrained
by the X-ray emission profile (\citealt{Ota2012}) and additional multiply imaged background
galaxies of differing redshifts from the quasar (\citealt{Oguri2008,Oguri2012b}), cluster lenses have
the potential of being excellent cosmological probes if it can be
demonstrated that the effects of substructure are controllable.
Here we present a light curve for the brighter A and B images of SDSS~J1029+2623 
spanning six 8-month observing seasons and measure their time delay. Because image C is faint and very
close to B, we cannot independently determine
its delay given the quality of our images.  Section \ref{sec:observations} summarizes the available data, Section \ref{sec:delay} derives the time
delay, and Section \ref{sec:discussion} discusses the results.

\section{Observations}
\label{sec:observations}

\begin{figure}[t]
\epsscale{1.0}
\plotone{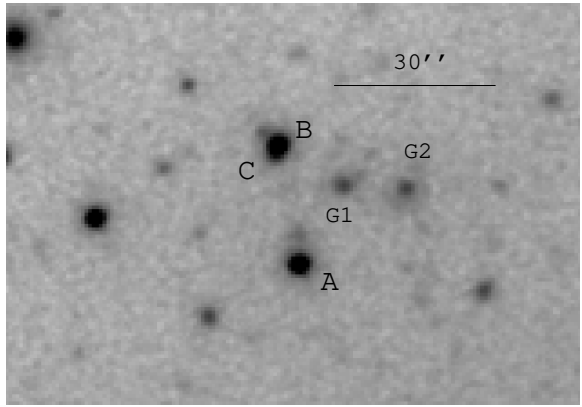}
\caption{SDSS \textit{r}-band Keplercam reference image of SDSS J1029+2623.
  This image combines 26 of the highest resolution epochs, for a total exposure
  time of 130 minutes.  
  One can see that the shape of the merged images B and C is different from A.\label{image}}
\end{figure}

We have monitored SDSS~J1029+2623 using Keplercam at the 
Fred Lawrence Whipple Observatory (FLWO) 1.2m telescope
 over a period of 1960 days from 
2007 January to 2012 June with an average sampling rate 
of three times a week when the source was visible.  
Keplercam has 0\farcs672
pixels, and the 1.2m telescope presently delivers very
poor quality images due to problems with its primary 
mirror.  While the data are adequate for monitoring the
widely separated A and B images, our light curve of image
B is that of B and C combined.  Since image C is relatively
faint ($\hbox{\textit{R}}\sim 20.3$) and expected to have a very short delay relative to
image B ($2$--$3$~days, see Section \ref{sec:discussion}), merging 
its flux with that of image B will have no significant consequences 
for our present results.   Each epoch consisted of three 5 minute exposures in the {\textit{r} filter.
The data were reduced using standard methods.

Although we analyzed the data following the methods used previously
for the quintuple quasar SDSS~J1004+4112 (\citealt{Fohlmeister2007}, \citealt{Fohlmeister2008})
and found consistent results, here we use
the ISIS difference imaging package 
(\citealt{Alard1998}).  The reference image shown in
Figure~\ref{image} was constructed from 26 of the best quality
sub-images, corresponding to a total integration time of
130~minutes.  The quasar images are labeled A and B following 
the notation of \cite{Inada2006}.  The third quasar image C lies roughly 
1\farcs8 south of B (\citealt{Oguri2008}). While it is marginally resolved 
in the reference image, we cannot obtain a reliable, independent light curve
for it.  The differential light curves of A and B+C were extracted 
following the standard procedures for ISIS.  

To calibrate the 
reference image, we matched 39 sources in the field to stars
with $15 < r < 21$ in the Sloan Digital Sky Survey (SDSS) DR8 (\citealt{Aihara2011}) catalogs.  We defined
the zero points using the SDSS magnitudes and Sextractor 
(\citealt{Bertin1996}) 5\farcs4 (8 pixel) diameter aperture magnitudes for sources
on the reference image.  After dropping objects more than 0.1~mag
from the median of the initial individual zero point estimates,
the nominal uncertainties in the calibration are negligible 
(2~mmag).  ISIS tends to modestly underestimate photometric
errors. Using a combination of the light curves of stars of magnitudes
comparable to those of the quasars and the statistics of points in the quasar
light curves with small temporal separations,\footnote{We take three points
separated by less than 7 days, predict the value of the middle points by
linearly interpolating the outer points and compare the difference between
the observed and predicted values of the middle point to the estimated
photometric errors.} we estimate that we 
must rescale the ISIS errors by a factor of $1.24$.
Combining the reference image photometry with 
the difference imaging light curves, we obtain the calibrated
light curves presented in Table~\ref{tab:lc}, where we report
and use the rescaled error bars.      
Figure~\ref{fig:lc1} shows the light curves. Here we have added as a 
first epoch the original SDSS observations of the system (\citealt{Inada2006}).

\begin{figure*}[t]
\epsscale{1.60}
\plotone{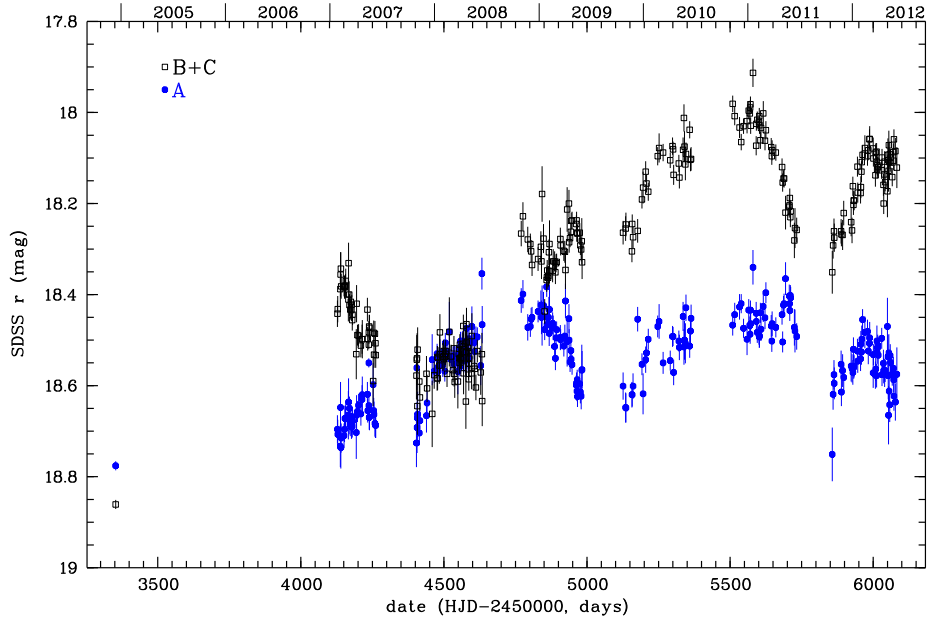}
\caption{Optical \textit{r}-band light curves for image A (filled blue circles) and images B$+$C (black squares). 
The first data point for both light curves is from the original SDSS observations.   
    }
\label{fig:lc1}
\end{figure*}

\begin{figure}[p]
\epsscale{.80}
 \plotone{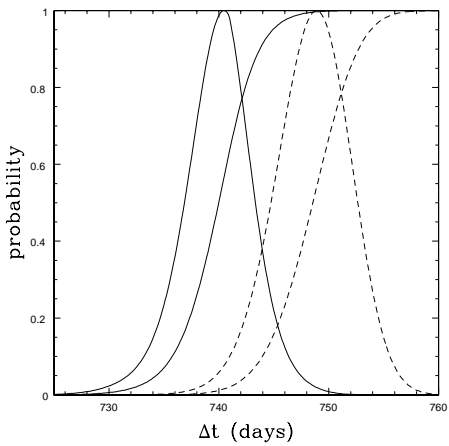}
\caption{Polynomial method probability distributions for the time delay based on either the AIC (solid) or BIC (dashed)
  information criteria for combining the models.  Both the differential and integral probability distributions are shown,
  with the differential probability normalized to its maximum. 
  }
  \label{fig:delay}
\end{figure}

\begin{figure*}[t]
\epsscale{1.60}
\plotone{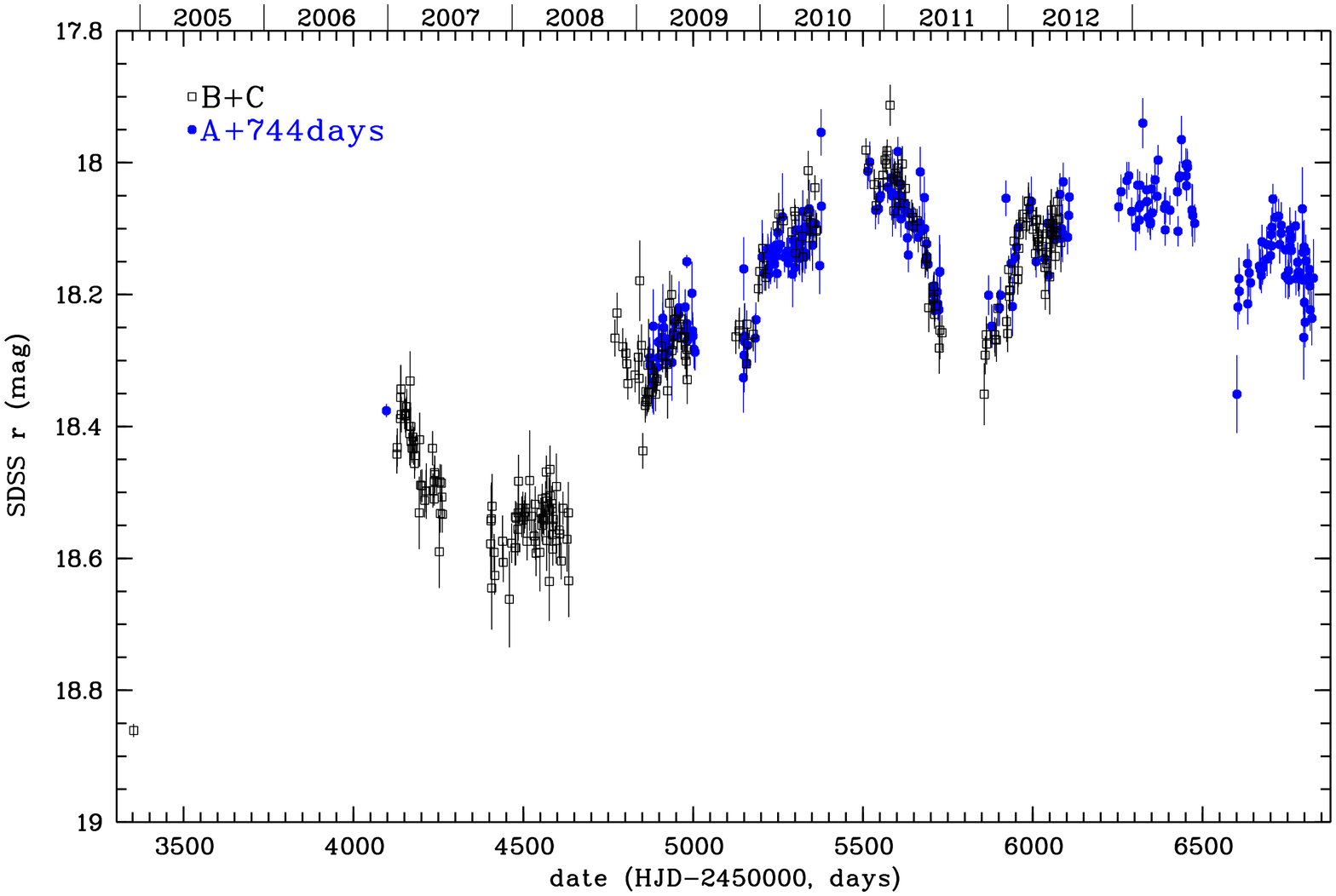}
\caption{Light curves for image A (filled blue circles) and images B+C (black squares) after shifting image A
     by the estimated time delay of $\Delta t_{AB} = (744 \pm 10)$~days and $-0.40$~mag.
     While not included in the models, note that the data point for A from the original SDSS 
     observations, shown by the blue circle at the start of the light curves, 
     matches well the start of our light curve for image B$+$C.  
    }
\label{fig:lc}
\end{figure*}

\section{Time Delay}
\label{sec:delay}

The challenge in measuring time delays is that the final uncertainties in essence depend on the nature
of the interpolation of the light curves used for the comparison \citep{kundic1997}.  Experience demonstrates
that it is worth considering multiple methods and that the formal uncertainties are typically
smaller than the actual uncertainties (when tested by improved light curves; e.g., \citealt{kochanek2}
and \citealt{Courbin2011} in the case of HE~0435$-$1223).  These issues will be most problematic
for short delays of the order of a few days both because the delay is not that different from the sampling cadence and
because quasars show less and less variability power on shorter time scales (e.g. \citealt{Macleod2010}).
There is also a cosmic variance of several percent in time delays produced by 
fluctuations in the mean density along the line of sight (e.g. \citealt{Barkana1996}, \citealt{Wambs2005}), although
one can attempt to use the visible galaxies in the field to estimate its amplitude 
(e.g. \citealt{Suyu2010}).  Delay ratios do benefit from higher accuracy measurements because they
are little affected by this cosmic variance.  Because cluster lenses like SDSS~J1029$+$2623 have 
relatively long delays ($\sim\hbox{years}$), it is easier to measure cosmic variance-limited delays than for 
single galaxy lenses.

Here we determine the time delays between images A and the combined B$+$C image pair,
where we know from simply shifting the light curves by hand that the delay is on the
order of two years.  We first consider two methods that do not directly test for 
microlensing or model its effects. If we first consider the simple $\chi^2$-minimization 
method described in \cite{Fohlmeister2007}, we measure a delay of $\Delta t_{AB}=(746 \pm 6)$ days (all delays
are in the sense of A leading B). The dispersion method of \cite{pelt1,pelt2}
gives a delay of $\Delta t_{AB} = (745 \pm 10)$~days.
The \cite{kochanek2} polynomial method,
where the source light curve and microlensing magnifications are described by polynomials, lets us examine the effects of microlensing.
We modeled each season with polynomials of all orders from $N_s=3$ to $25$ and with 
constant, linear or quadratic ($N_\mu=0$ to $2$) polynomials for the microlensing variability in each 
season.  

Models allowing for no microlensing parameters were strongly ruled out
by the data.
We evaluate the models using Bayesian information criteria to weight the changes 
in the numbers of parameters between models, where the
probability of a delay with a goodness of fit $\chi^2$ is $\exp(-\chi^2/2 - k N_p)$ 
where $N_p=4(N_s+N_\mu)$ is the number of parameters in the model.\footnote{Each of the four ``seasons'' is described
by a source polynomial of order $N_s$ and a microlensing polynomial of order $N_\mu$.} 
We consider both the more liberal Akaike information criterion (AIC)
with $k=1$, and the more conservative Bayesian information criterion (BIC) with $k=\ln N_{data}$
(see \citealt{Poindexter2007}), where $N_{data} = 355$ is the number of data points in the 
seasons that overlap.  The
AIC favors $N_s \sim 7$ and $N_\mu=1$ (formally, the relative probabilities for $N_\mu=0$, $1$ 
and $2$ are
$0.19:1.00:0.11$), while the BIC favors $N_s \simeq 5 $ and $N_\mu =0$ ($1.0:0.0:0.0$). 
The combined result for the AIC is a median time delay of $\Delta t_{AB} = 740.0$ with a
90\% confidence range of $739.0 < \Delta t_{AB} < 744.9$, while that for the BIC is a
median of $\Delta t_{AB} = 748.6$ with a 90\% confidence range of $742.6 < \Delta t_{AB} < 753.8$.
As Figure~\ref{fig:delay} shows, it is difficult to evaluate the relative merits
of these solutions by eye, although the low order polynomials favored by the BIC
seem to overly smooth the light curves independent of their statistical merits. If
we apply no information criterion at all, the median is $t_{AB}=739.3$ with a 90\%
confidence range of $730.8 < t_{AB} < 744.5$.

If we consider the two most probable models, $N_s=7$ and $N_\mu=1$ for the AIC and 
$N_s=5$ and $N_\mu=0$ for the BIC, we find that the effects of microlensing are small but
 statistically significant.  In the best BIC model, 
the flux ratios in the four seasons are $\Delta m_{AB}=-0.373\pm0.006$, $-0.413\pm0.006$, 
$-0.424\pm0.005$ and $-0.384\pm0.006$~mag for the four overlapping seasons. The
small fluctuations suggest the presence of microlensing, with the middle two seasons
showing a significant shift of about 0.05~mag relative to the first and last seasons.  
The AIC model shifts are very similar but have marginally detected gradients of 
$-0.04\pm0.06$, $-0.03\pm0.03$, $0.08\pm0.03$ and $-0.07\pm0.04$ ~mag/yr$^{-1}$.  Given
the quality of the data, these effects are not obvious, and it is not
inconceivable that they are partly due to systematics from matching data taken
at very different epochs even though ISIS excels at properly cross calibrating
data and correcting for point-spread function differences.  
In fact, when we fit the light curves of stars with similar magnitudes, there were also
shifts of this amplitude.  Thus, while the addition of microlensing parameters leads to
statistically significant improvements in the fits, they could be modeling effects other
than microlensing. Given the low
amplitudes and the quality of the data we regard the evidence for
the detection of microlensing (as a physical process rather than as
parameters in a fitting function) to be marginal.

In the following we adopt a time delay of $\Delta t_{AB} = (744 \pm 10)$~days 
that conservatively combines the AIC and BIC results from the polynomial method.
Note, however, that this estimate is also in agreement with the $\chi^2$-minimization and the \cite{pelt1, pelt2} dispersion method results 
that do not take possible microlensing variations into account.
Figure \ref{fig:lc} shows the lightcurves of the A and B$+$C images shifted by the determined 
time delay of $\Delta t_{AB} = (744 \pm 10)$~days and $-0.40$~mag. The shifted photometry for image A
from the original SDSS observations in 2004 closely matches the
start of our light curves although it is not quite overlapping.  The later \textit{R}-band 
observations by \cite{Oguri2008} are contemporaneous with the start of our observations,
so we lack any additional data in the gap between 2004 and 2007.

\begin{figure}[p]
\epsscale{1.00}
\plotone{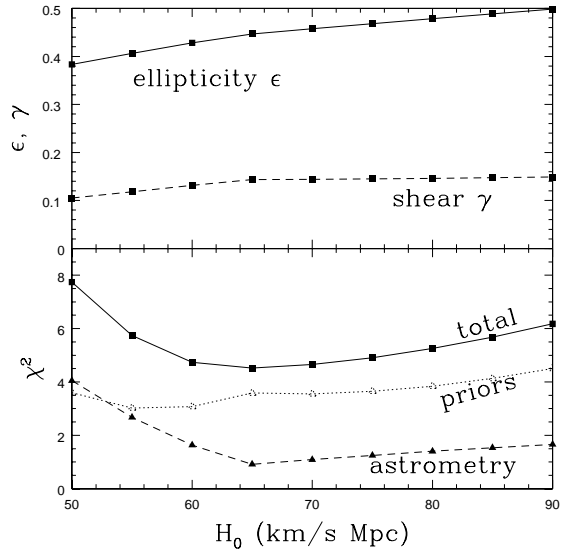}
\caption{Goodness of fit as a function of the Hubble constant $H_0$ for simple
     models consisting of an ellipsoidal NFW model in an external shear.  The
     solid line (black squares) show the total $\chi^2$ statistic for the
     goodness of fit, which is dominated by contributions from the image
     positions (astrometry, filled triangles, dashed curve) and the 
     ellipticity and lens position priors (priors, open triangles, dotted curves)
     with negligible contributions from matching the time delays. The
     upper panels show the trend of the models towards flatter density
     distributions, where the axis ratio of the model is $1-\epsilon$, and
     higher external shear $\gamma$ for increasing $H_0$.
     }
  \label{fig:models}
\end{figure}

 The measured time delay allows to assemble a densely sampled
light curve spanning more than 1000 days in rest frame. Unfortunately, the merged light curves 
do not fill in the seasonal gaps due to the delay of almost exactly 
two years (see Figure~\ref{fig:lc}). We computed the structure function, which represents the magnitude change as a function of  time lag between two photometric measurements. The intrinsic variability from the merged A/B+C light curves of this distant individual quasar shows a power-law slope of $0.32\pm0.02$ and a variability amplitude at 100 days of $0.15\pm0.03$ mag.

\section{Summary and Discussion}
\label{sec:discussion}

We presented 5.4 years of optical monitoring for the two bright
lensed quasar images of the largest image separation 
 gravitationally lensed quasar SDSS J1029+2623. 
 We find that image A leads the images B and C by 
 $\Delta t_{AB} = (744 \pm 10)$~days (90\% confidence). The formal error bar on the time delay, which includes 
 statistical uncertainties and systematic differences,
 is $\sim 1.3\%$ and is in
 the regime where cosmic variance caused by fluctuations in
 the mean line-of-sight density is more important than the measurement errors. 
 We find that the effect of microlensing in this system is small. Formally, the detection is
   statistically significant, but we view the overall evidence for microlensing rather than
   low level systematic uncertainties as weak. 
   This is the second longest measured time delay after
 the 822 day delay between images C and A in SDSS~J1004+4112 (\citealt{Fohlmeister2008}).
 
A detailed interpretation of the measured delay is deferred pending the completion of our
analysis of the \textit{HST} images (\citealt{Oguri2012b}) and additional spectroscopy of the lensed
arcs in this system.  However, as an experiment, we fit the lens using a Navarro-Frenk-White model centered 
near galaxy G2 with a break radius of 78\farcs0 based on the X-ray data (\citealt{Ota2012}). 
We adopt the component positions ($\pm 0\farcs05$) from \cite{Oguri2008} and a time delay of 
$\Delta t_{AB} = (744 \pm 10)$~days that tries to conservatively combine the AIC and BIC results.
We used priors of $\pm 1\farcs0$ on 
the position of the model relative to galaxy G2,
$\epsilon = 0.46\pm0.05$ for the ellipticity of the cluster density, and
$\gamma = 0.05 \pm 0.05$ for any additional external shear.  The value of
$\epsilon$ was determined by the best fit model with $H_0=70$~km~s$^{-1}$~Mpc$^{-1}$.
We then fit models as a function of the Hubble constant $H_0$ assuming a flat, $\Omega_0=0.3$,
and $\Lambda_0=0.7$ cosmological model.   We did not include the flux ratios 
in the fits because of the known flux ratio anomaly.  The anomaly appears to
be due to a small galaxy near image C (\citealt{Oguri2012b}) which should have
little effect on the overall geometry and the AB time delay.  
In a normal four-image lens, the constraint on the radial mass profile from the X-ray data would largely 
eliminate any degeneracies because they are created by uncertainties in the
surface mass density at the radius of the images (\citealt{kochanek2002}).  

Figure~\ref{fig:models} shows the resulting goodness of fit as a function of $H_0$ for 
these simple models.  Low values of $H_0$ are disfavored by the astrometry, while high
values are weakly disfavored by the priors on the shape of the potential.  There is
clearly a strong need for additional constraints in order to use this delay for the full characterization of the lens system.  There
is certainly no difficulty improving the astrometric constraints, since with \textit{HST} it will be 
trivial to increase the astrometric precision from the $0\farcs05$ used here to $0\farcs01$ 
or smaller, although the concern here will be whether noise from unmodeled sub-structures 
in the cluster dominate over the measurement precision.  More important, however,
is the addition of strong constraints on the shape of the potential, since there
is a strong trend requiring flatter density distributions and higher external 
shears for larger values of $H_0$.  We already know from \cite{Oguri2008}
that there is a lensed image of the quasar host galaxy as well as several additional
arc systems, and this is confirmed by the new \textit{HST} observations (\citealt{Oguri2012b}). 
The primary difference between these models and the early model of \cite{Oguri2008}, which 
predicted a longer delay, is that these models are centered on G2 rather than on G1 based on the X-ray data.  
 Particularly
if combined with additional arc redshifts and direct mass models of the X-ray emission,
there are no obvious barriers to greatly improving on Figure~\ref{fig:models}.  Where
\cite{Suyu2010} use dynamical measurements of the central lens galaxy to break the
\cite{kochanek2002} or other degeneracies, here we should be able to do so using
a combination of the X-ray and arc data.  There is, however, significant evidence
that the cluster is undergoing a merger, which means that the additional lensing
constraints will be more reliable constraints on the mass distribution than further
X-ray observations.  On the other hand, the ability to combine deeper X-ray observations with the
lensing constraints makes this system and excellent laboratory for studying the
effects of mergers on the X-ray properties of clusters.

In our present data, measuring the B-C time delay is impossible given the data
quality.  It does shift steadily over the model sequence, decreasing with 
increasing $H_0$ from 3.4 to 2.1 days.  This fully justifies our making the
measurements using the combined B/C light curve.  It is well worth measuring
the BC delay because it is very sensitive to the perturbing galaxy that
causes the flux ratio anomaly (see \citealt{Oguri2007}, \citealt{Keeton2009})
and hence probes the structure of galaxy halos in cluster environments.  
Such short delays are extremely difficult to measure accurately 
in ground based observations because quasars
have so little variability power on such short time scales 
(see \citealt{Mushotzky2011}) and might require a space-based lens
monitoring satellite like the proposed OMEGA (\citealt{Moustakas2008}).

\acknowledgments

We thank all the participating observers at the Harvard-Smithsonian Center for Astrophysics 
for their support of these observations.
Our observations were obtained with the F. L. Whipple 1.2m telescope, with support from
the Smithsonian Astrophysical Observatory. 
CSK is supported by NSF grant AST-1009756.
This work was supported in part by the FIRST program
``Subaru Measurements of Images and Redshifts (SuMIRe)'', World Premier
International Research Center Initiative (WPI Initiative), MEXT,
Japan, and Grant-in-Aid for Scientific Research from the JSPS
(23740161). 
Funding for SDSS-III has been provided by the Alfred P. Sloan Foundation, the Participating Institutions, the National Science Foundation, and the U.S. Department of Energy Office of Science. The SDSS-III web site is http://www.sdss3.org/.
SDSS-III is managed by the Astrophysical Research Consortium for the Participating Institutions of the SDSS-III Collaboration including the University of Arizona, the Brazilian Participation Group, Brookhaven National Laboratory, University of Cambridge, Carnegie Mellon University, University of Florida, the French Participation Group, the German Participation Group, Harvard University, the Instituto de Astrofisica de Canarias, the Michigan State/Notre Dame/JINA Participation Group, Johns Hopkins University, Lawrence Berkeley National Laboratory, Max Planck Institute for Astrophysics, New Mexico State University, New York University, Ohio State University, Pennsylvania State University, University of Portsmouth, Princeton University, the Spanish Participation Group, University of Tokyo, University of Utah, Vanderbilt University, University of Virginia, University of Washington, and Yale University.

\begin{deluxetable}{lcc}
\tabletypesize{\scriptsize}
\tablecaption{Light Curves }
\tablewidth{0pt}
\tablehead{
\multicolumn{1}{c}{HJD-2450000} &
\multicolumn{1}{c}{A (mag)}  &
\multicolumn{1}{c}{B(+C) (mag)} }
\startdata
$3352.900$ &$18.776 \pm  0.010$ &$18.861 \pm  0.010$ \\ 
$4127.878$ &$18.696 \pm  0.031$ &$18.442 \pm  0.029$ \\ 
$4128.910$ &$18.707 \pm  0.032$ &$18.432 \pm  0.029$ \\ 
$4137.945$ &$18.733 \pm  0.046$ &$18.388 \pm  0.038$ \\ 
$4138.794$ &$18.648 \pm  0.056$ &$18.356 \pm  0.049$ \\ 
$4139.819$ &$18.736 \pm  0.046$ &$18.343 \pm  0.036$ \\ 
$4140.835$ &$18.715 \pm  0.032$ &$18.382 \pm  0.027$ \\ 
$4150.891$ &$18.710 \pm  0.027$ &$18.381 \pm  0.023$ \\ 
$4152.773$ &$18.696 \pm  0.028$ &$18.384 \pm  0.025$ \\ 
$4153.772$ &$18.672 \pm  0.028$ &$18.383 \pm  0.025$ \\ 
$4155.860$ &$18.672 \pm  0.028$ &$18.380 \pm  0.025$ \\ 
$4156.841$ &$18.671 \pm  0.032$ &$18.370 \pm  0.027$ \\ 
$4165.937$ &$18.665 \pm  0.053$ &$18.411 \pm  0.048$ \\ 
$4166.780$ &$18.636 \pm  0.052$ &$18.331 \pm  0.045$ \\ 
$4168.873$ &$18.675 \pm  0.032$ &$18.400 \pm  0.029$ \\ 
$4169.826$ &$18.650 \pm  0.026$ &$18.423 \pm  0.024$ \\ 
$4172.730$ &$18.687 \pm  0.024$ &$18.433 \pm  0.022$ \\ 
$4174.794$ &$18.667 \pm  0.025$ &$18.416 \pm  0.023$ \\ 
$4176.867$ &$18.691 \pm  0.026$ &$18.434 \pm  0.024$ \\ 
$4177.805$ &$18.685 \pm  0.023$ &$18.422 \pm  0.021$ \\ 
$4179.682$ &$18.689 \pm  0.025$ &$18.456 \pm  0.023$ \\ 
$4180.702$ &$18.678 \pm  0.025$ &$18.444 \pm  0.023$ \\ 
$4193.775$ &$18.703 \pm  0.058$ &$18.531 \pm  0.055$ \\ 
$4194.814$ &$18.665 \pm  0.044$ &$18.420 \pm  0.041$ \\ 
$4197.741$ &$18.657 \pm  0.023$ &$18.489 \pm  0.023$ \\ 
$4201.799$ &$18.643 \pm  0.025$ &$18.490 \pm  0.025$ \\ 
$4209.686$ &$18.662 \pm  0.026$ &$18.512 \pm  0.026$ \\ 
$4213.818$ &$18.630 \pm  0.031$ &$18.498 \pm  0.032$ \\ 
$4214.667$ &$18.620 \pm  0.040$ &$18.498 \pm  0.042$ \\ 
$4232.721$ &$18.619 \pm  0.027$ &$18.433 \pm  0.026$ \\ 
$4233.753$ &$18.655 \pm  0.026$ &$18.497 \pm  0.026$ \\ 
$4237.500$ &$18.550 \pm  0.010$ &$18.510 \pm  0.010$ \\ 
$4237.688$ &$18.644 \pm  0.025$ &$18.483 \pm  0.025$ \\ 
$4238.703$ &$18.670 \pm  0.027$ &$18.470 \pm  0.026$ \\ 
$4239.760$ &$18.669 \pm  0.029$ &$18.473 \pm  0.028$ \\ 
$4252.658$ &$18.598 \pm  0.048$ &$18.590 \pm  0.055$ \\ 
$4254.675$ &$18.655 \pm  0.028$ &$18.484 \pm  0.027$ \\ 
$4255.683$ &$18.663 \pm  0.028$ &$18.532 \pm  0.029$ \\ 
$4258.690$ &$18.683 \pm  0.029$ &$18.486 \pm  0.028$ \\ 
$4260.712$ &$18.687 \pm  0.028$ &$18.507 \pm  0.027$ \\ 
$4261.698$ &$18.687 \pm  0.028$ &$18.533 \pm  0.028$ \\ 
$4404.003$ &$18.726 \pm  0.053$ &$18.578 \pm  0.052$ \\ 
$4405.003$ &$18.561 \pm  0.048$ &$18.543 \pm  0.054$ \\ 
$4405.981$ &$18.692 \pm  0.056$ &$18.540 \pm  0.055$ \\ 
$4407.006$ &$18.671 \pm  0.057$ &$18.645 \pm  0.063$ \\ 
$4407.993$ &$18.663 \pm  0.050$ &$18.521 \pm  0.049$ \\ 
$4414.021$ &$18.704 \pm  0.027$ &$18.591 \pm  0.028$ \\ 
$4416.001$ &$18.677 \pm  0.027$ &$18.626 \pm  0.029$ \\ 
$4439.043$ &$18.666 \pm  0.037$ &$18.574 \pm  0.039$ \\ 
$4440.999$ &$18.638 \pm  0.026$ &$18.606 \pm  0.030$ \\ 
$4459.013$ &$18.543 \pm  0.056$ &$18.662 \pm  0.073$ \\ 
$4466.002$ &$18.568 \pm  0.026$ &$18.577 \pm  0.030$ \\ 
$4475.979$ &$18.530 \pm  0.022$ &$18.584 \pm  0.027$ \\ 
$4477.013$ &$18.534 \pm  0.021$ &$18.537 \pm  0.024$ \\ 
$4477.955$ &$18.560 \pm  0.022$ &$18.584 \pm  0.026$ \\ 
$4478.977$ &$18.569 \pm  0.022$ &$18.539 \pm  0.024$ \\ 
$4483.849$ &$18.550 \pm  0.034$ &$18.556 \pm  0.040$ \\ 
$4484.941$ &$18.542 \pm  0.037$ &$18.531 \pm  0.042$ \\ 
$4485.951$ &$18.535 \pm  0.037$ &$18.483 \pm  0.040$ \\ 
$4495.906$ &$18.554 \pm  0.027$ &$18.537 \pm  0.031$ \\ 
$4498.900$ &$18.542 \pm  0.022$ &$18.524 \pm  0.025$ \\ 
$4499.883$ &$18.525 \pm  0.022$ &$18.542 \pm  0.025$ \\ 
$4502.989$ &$18.568 \pm  0.023$ &$18.537 \pm  0.025$ \\ 
$4504.919$ &$18.506 \pm  0.022$ &$18.530 \pm  0.025$ \\ 
$4505.958$ &$18.532 \pm  0.021$ &$18.524 \pm  0.024$ \\ 
$4510.952$ &$18.524 \pm  0.024$ &$18.574 \pm  0.029$ \\ 
$4518.867$ &$18.482 \pm  0.066$ &$18.482 \pm  0.076$ \\ 
$4524.882$ &$18.543 \pm  0.023$ &$18.536 \pm  0.026$ \\ 
$4531.700$ &$18.536 \pm  0.022$ &$18.566 \pm  0.026$ \\ 
$4534.946$ &$18.534 \pm  0.023$ &$18.518 \pm  0.027$ \\ 
$4535.786$ &$18.552 \pm  0.023$ &$18.574 \pm  0.028$ \\ 
$4537.630$ &$18.539 \pm  0.028$ &$18.592 \pm  0.035$ \\ 
$4548.789$ &$18.569 \pm  0.050$ &$18.591 \pm  0.059$ \\ 
$4551.717$ &$18.518 \pm  0.020$ &$18.531 \pm  0.023$ \\ 
$4553.846$ &$18.544 \pm  0.023$ &$18.540 \pm  0.027$ \\ 
$4554.645$ &$18.526 \pm  0.020$ &$18.550 \pm  0.024$ \\ 
$4555.642$ &$18.559 \pm  0.021$ &$18.510 \pm  0.023$ \\ 
$4557.804$ &$18.503 \pm  0.021$ &$18.539 \pm  0.024$ \\ 
$4558.726$ &$18.502 \pm  0.020$ &$18.539 \pm  0.023$ \\ 
$4560.751$ &$18.532 \pm  0.020$ &$18.542 \pm  0.023$ \\ 
$4565.831$ &$18.502 \pm  0.022$ &$18.508 \pm  0.025$ \\ 
$4567.789$ &$18.545 \pm  0.024$ &$18.469 \pm  0.025$ \\ 
$4568.779$ &$18.523 \pm  0.037$ &$18.573 \pm  0.046$ \\ 
$4569.707$ &$18.535 \pm  0.036$ &$18.513 \pm  0.042$ \\ 
$4575.724$ &$18.519 \pm  0.044$ &$18.527 \pm  0.050$ \\ 
$4576.730$ &$18.534 \pm  0.047$ &$18.635 \pm  0.060$ \\ 
$4577.771$ &$18.511 \pm  0.040$ &$18.526 \pm  0.045$ \\ 
$4578.768$ &$18.474 \pm  0.032$ &$18.465 \pm  0.036$ \\ 
$4579.698$ &$18.541 \pm  0.022$ &$18.505 \pm  0.024$ \\ 
$4583.680$ &$18.499 \pm  0.023$ &$18.517 \pm  0.027$ \\ 
$4584.703$ &$18.543 \pm  0.025$ &$18.524 \pm  0.029$ \\ 
$4585.747$ &$18.537 \pm  0.021$ &$18.564 \pm  0.024$ \\ 
$4586.699$ &$18.539 \pm  0.021$ &$18.586 \pm  0.025$ \\ 
$4587.683$ &$18.497 \pm  0.021$ &$18.551 \pm  0.025$ \\ 
$4588.648$ &$18.540 \pm  0.021$ &$18.541 \pm  0.024$ \\ 
$4596.739$ &$18.504 \pm  0.029$ &$18.573 \pm  0.035$ \\ 
$4597.703$ &$18.470 \pm  0.044$ &$18.491 \pm  0.050$ \\ 
$4605.654$ &$18.505 \pm  0.036$ &$18.557 \pm  0.043$ \\ 
$4607.679$ &$18.525 \pm  0.039$ &$18.563 \pm  0.046$ \\ 
$4611.712$ &$18.492 \pm  0.022$ &$18.604 \pm  0.028$ \\ 
$4616.671$ &$18.493 \pm  0.021$ &$18.524 \pm  0.025$ \\ 
$4628.648$ &$18.556 \pm  0.043$ &$18.571 \pm  0.049$ \\ 
$4632.651$ &$18.354 \pm  0.035$ &$18.531 \pm  0.047$ \\ 
$4633.655$ &$18.466 \pm  0.041$ &$18.634 \pm  0.055$ \\ 
$4769.958$ &$18.413 \pm  0.027$ &$18.266 \pm  0.028$ \\ 
$4776.021$ &$18.399 \pm  0.031$ &$18.228 \pm  0.031$ \\ 
$4793.023$ &$18.472 \pm  0.029$ &$18.279 \pm  0.028$ \\ 
$4801.948$ &$18.470 \pm  0.023$ &$18.289 \pm  0.023$ \\ 
$4804.847$ &$18.454 \pm  0.025$ &$18.305 \pm  0.025$ \\ 
$4807.942$ &$18.450 \pm  0.023$ &$18.335 \pm  0.024$ \\ 
$4829.042$ &$18.437 \pm  0.024$ &$18.322 \pm  0.026$ \\ 
$4838.864$ &$18.423 \pm  0.032$ &$18.295 \pm  0.034$ \\ 
$4840.957$ &$18.450 \pm  0.037$ &$18.327 \pm  0.039$ \\ 
$4842.930$ &$18.446 \pm  0.066$ &$18.179 \pm  0.061$ \\ 
$4847.995$ &$18.424 \pm  0.030$ &$18.277 \pm  0.032$ \\ 
$4851.991$ &$18.477 \pm  0.023$ &$18.437 \pm  0.027$ \\ 
$4859.020$ &$18.383 \pm  0.022$ &$18.368 \pm  0.026$ \\ 
$4859.999$ &$18.466 \pm  0.024$ &$18.362 \pm  0.026$ \\ 
$4861.001$ &$18.450 \pm  0.022$ &$18.347 \pm  0.023$ \\ 
$4863.932$ &$18.450 \pm  0.019$ &$18.363 \pm  0.021$ \\ 
$4864.895$ &$18.450 \pm  0.021$ &$18.359 \pm  0.022$ \\ 
$4866.841$ &$18.432 \pm  0.027$ &$18.307 \pm  0.028$ \\ 
$4867.868$ &$18.485 \pm  0.028$ &$18.349 \pm  0.029$ \\ 
$4868.854$ &$18.433 \pm  0.051$ &$18.289 \pm  0.052$ \\ 
$4879.646$ &$18.463 \pm  0.024$ &$18.348 \pm  0.025$ \\ 
$4882.028$ &$18.473 \pm  0.027$ &$18.327 \pm  0.027$ \\ 
$4886.686$ &$18.514 \pm  0.023$ &$18.327 \pm  0.022$ \\ 
$4889.620$ &$18.540 \pm  0.026$ &$18.351 \pm  0.026$ \\ 
$4891.858$ &$18.495 \pm  0.020$ &$18.331 \pm  0.020$ \\ 
$4893.874$ &$18.477 \pm  0.018$ &$18.328 \pm  0.018$ \\ 
$4906.858$ &$18.494 \pm  0.027$ &$18.278 \pm  0.025$ \\ 
$4907.657$ &$18.499 \pm  0.023$ &$18.292 \pm  0.022$ \\ 
$4918.854$ &$18.513 \pm  0.023$ &$18.304 \pm  0.022$ \\ 
$4922.675$ &$18.491 \pm  0.025$ &$18.306 \pm  0.024$ \\ 
$4924.785$ &$18.414 \pm  0.038$ &$18.346 \pm  0.042$ \\ 
$4930.757$ &$18.503 \pm  0.055$ &$18.213 \pm  0.049$ \\ 
$4936.810$ &$18.453 \pm  0.032$ &$18.200 \pm  0.030$ \\ 
$4937.655$ &$18.500 \pm  0.024$ &$18.285 \pm  0.023$ \\ 
$4940.663$ &$18.548 \pm  0.022$ &$18.276 \pm  0.020$ \\ 
$4943.649$ &$18.523 \pm  0.021$ &$18.262 \pm  0.020$ \\ 
$4944.683$ &$18.542 \pm  0.021$ &$18.238 \pm  0.018$ \\ 
$4947.703$ &$18.554 \pm  0.022$ &$18.237 \pm  0.019$ \\ 
$4962.657$ &$18.599 \pm  0.022$ &$18.245 \pm  0.019$ \\ 
$4963.658$ &$18.594 \pm  0.023$ &$18.237 \pm  0.019$ \\ 
$4964.670$ &$18.612 \pm  0.022$ &$18.265 \pm  0.019$ \\ 
$4965.672$ &$18.587 \pm  0.022$ &$18.264 \pm  0.019$ \\ 
$4966.675$ &$18.624 \pm  0.023$ &$18.266 \pm  0.019$ \\ 
$4974.679$ &$18.596 \pm  0.025$ &$18.264 \pm  0.021$ \\ 
$4976.685$ &$18.614 \pm  0.025$ &$18.292 \pm  0.021$ \\ 
$4979.673$ &$18.623 \pm  0.029$ &$18.301 \pm  0.026$ \\ 
$4981.679$ &$18.566 \pm  0.056$ &$18.283 \pm  0.051$ \\ 
$4982.720$ &$18.565 \pm  0.041$ &$18.329 \pm  0.037$ \\ 
$5125.985$ &$18.601 \pm  0.030$ &$18.264 \pm  0.026$ \\ 
$5135.020$ &$18.648 \pm  0.033$ &$18.255 \pm  0.027$ \\ 
$5136.013$ &$18.649 \pm  0.032$ &$18.246 \pm  0.026$ \\ 
$5157.029$ &$18.620 \pm  0.028$ &$18.305 \pm  0.024$ \\ 
$5157.997$ &$18.620 \pm  0.026$ &$18.245 \pm  0.021$ \\ 
$5160.956$ &$18.601 \pm  0.028$ &$18.274 \pm  0.024$ \\ 
$5177.026$ &$18.454 \pm  0.027$ &$18.260 \pm  0.026$ \\ 
$5192.028$ &$18.553 \pm  0.024$ &$18.191 \pm  0.020$ \\ 
$5195.977$ &$18.618 \pm  0.045$ &$18.165 \pm  0.035$ \\ 
$5204.997$ &$18.543 \pm  0.029$ &$18.130 \pm  0.023$ \\ 
$5208.020$ &$18.528 \pm  0.024$ &$18.156 \pm  0.020$ \\ 
$5214.035$ &$18.498 \pm  0.023$ &$18.174 \pm  0.020$ \\ 
$5246.959$ &$18.470 \pm  0.024$ &$18.096 \pm  0.020$ \\ 
$5251.857$ &$18.459 \pm  0.038$ &$18.078 \pm  0.032$ \\ 
$5265.907$ &$18.550 \pm  0.024$ &$18.088 \pm  0.019$ \\ 
$5290.785$ &$18.545 \pm  0.022$ &$18.105 \pm  0.017$ \\ 
$5298.764$ &$18.492 \pm  0.023$ &$18.074 \pm  0.019$ \\ 
$5300.794$ &$18.492 \pm  0.021$ &$18.081 \pm  0.017$ \\ 
$5302.645$ &$18.571 \pm  0.028$ &$18.137 \pm  0.022$ \\ 
$5320.794$ &$18.502 \pm  0.023$ &$18.112 \pm  0.019$ \\ 
$5322.797$ &$18.516 \pm  0.028$ &$18.143 \pm  0.024$ \\ 
$5335.696$ &$18.448 \pm  0.032$ &$18.082 \pm  0.027$ \\ 
$5338.680$ &$18.515 \pm  0.039$ &$18.012 \pm  0.030$ \\ 
$5340.669$ &$18.500 \pm  0.043$ &$18.075 \pm  0.034$ \\ 
$5343.675$ &$18.507 \pm  0.044$ &$18.114 \pm  0.035$ \\ 
$5345.657$ &$18.429 \pm  0.029$ &$18.091 \pm  0.025$ \\ 
$5358.654$ &$18.513 \pm  0.026$ &$18.038 \pm  0.019$ \\ 
$5361.665$ &$18.480 \pm  0.030$ &$18.104 \pm  0.025$ \\ 
$5363.681$ &$18.452 \pm  0.030$ &$18.102 \pm  0.026$ \\ 
$5509.008$ &$18.467 \pm  0.023$ &$17.981 \pm  0.018$ \\ 
$5516.029$ &$18.444 \pm  0.026$ &$18.008 \pm  0.020$ \\ 
$5533.044$ &$18.427 \pm  0.028$ &$18.033 \pm  0.023$ \\ 
$5538.857$ &$18.420 \pm  0.021$ &$18.065 \pm  0.018$ \\ 
$5547.002$ &$18.474 \pm  0.021$ &$18.030 \pm  0.017$ \\ 
$5559.060$ &$18.498 \pm  0.035$ &$18.019 \pm  0.027$ \\ 
$5565.041$ &$18.434 \pm  0.024$ &$17.996 \pm  0.019$ \\ 
$5569.048$ &$18.468 \pm  0.020$ &$18.002 \pm  0.015$ \\ 
$5570.034$ &$18.487 \pm  0.021$ &$17.989 \pm  0.015$ \\ 
$5571.054$ &$18.434 \pm  0.021$ &$17.982 \pm  0.017$ \\ 
$5572.017$ &$18.464 \pm  0.022$ &$18.029 \pm  0.018$ \\ 
$5580.062$ &$18.340 \pm  0.038$ &$17.913 \pm  0.031$ \\ 
$5591.053$ &$18.441 \pm  0.025$ &$18.073 \pm  0.021$ \\ 
$5591.810$ &$18.459 \pm  0.019$ &$18.025 \pm  0.015$ \\ 
$5594.990$ &$18.483 \pm  0.023$ &$18.016 \pm  0.018$ \\ 
$5600.907$ &$18.489 \pm  0.021$ &$18.021 \pm  0.016$ \\ 
$5602.016$ &$18.493 \pm  0.024$ &$18.008 \pm  0.018$ \\ 
$5602.742$ &$18.491 \pm  0.025$ &$18.028 \pm  0.019$ \\ 
$5605.987$ &$18.440 \pm  0.022$ &$18.061 \pm  0.019$ \\ 
$5608.030$ &$18.476 \pm  0.026$ &$18.034 \pm  0.020$ \\ 
$5616.008$ &$18.426 \pm  0.034$ &$18.002 \pm  0.027$ \\ 
$5621.979$ &$18.451 \pm  0.025$ &$18.062 \pm  0.021$ \\ 
$5624.972$ &$18.396 \pm  0.023$ &$18.039 \pm  0.020$ \\ 
$5643.645$ &$18.469 \pm  0.024$ &$18.076 \pm  0.020$ \\ 
$5645.839$ &$18.502 \pm  0.024$ &$18.096 \pm  0.019$ \\ 
$5646.917$ &$18.464 \pm  0.027$ &$18.083 \pm  0.022$ \\ 
$5659.832$ &$18.472 \pm  0.022$ &$18.088 \pm  0.018$ \\ 
$5681.640$ &$18.444 \pm  0.022$ &$18.120 \pm  0.019$ \\ 
$5683.704$ &$18.504 \pm  0.023$ &$18.154 \pm  0.020$ \\ 
$5686.721$ &$18.423 \pm  0.020$ &$18.145 \pm  0.018$ \\ 
$5688.639$ &$18.420 \pm  0.022$ &$18.145 \pm  0.020$ \\ 
$5693.706$ &$18.365 \pm  0.036$ &$18.220 \pm  0.037$ \\ 
$5706.738$ &$18.403 \pm  0.025$ &$18.207 \pm  0.024$ \\ 
$5707.682$ &$18.420 \pm  0.024$ &$18.204 \pm  0.022$ \\ 
$5708.680$ &$18.435 \pm  0.023$ &$18.188 \pm  0.021$ \\ 
$5710.662$ &$18.402 \pm  0.023$ &$18.230 \pm  0.022$ \\ 
$5711.660$ &$18.407 \pm  0.024$ &$18.217 \pm  0.024$ \\ 
$5724.679$ &$18.472 \pm  0.040$ &$18.281 \pm  0.039$ \\ 
$5726.666$ &$18.480 \pm  0.047$ &$18.254 \pm  0.044$ \\ 
$5732.664$ &$18.492 \pm  0.029$ &$18.258 \pm  0.027$ \\ 
$5857.014$ &$18.751 \pm  0.059$ &$18.351 \pm  0.047$ \\ 
$5860.003$ &$18.619 \pm  0.034$ &$18.292 \pm  0.029$ \\ 
$5863.007$ &$18.576 \pm  0.032$ &$18.261 \pm  0.028$ \\ 
$5863.982$ &$18.595 \pm  0.035$ &$18.275 \pm  0.031$ \\ 
$5887.995$ &$18.553 \pm  0.030$ &$18.261 \pm  0.027$ \\ 
$5889.027$ &$18.614 \pm  0.031$ &$18.268 \pm  0.026$ \\ 
$5893.027$ &$18.567 \pm  0.036$ &$18.269 \pm  0.033$ \\ 
$5897.037$ &$18.582 \pm  0.031$ &$18.221 \pm  0.027$ \\ 
$5923.052$ &$18.557 \pm  0.034$ &$18.241 \pm  0.031$ \\ 
$5926.044$ &$18.562 \pm  0.031$ &$18.259 \pm  0.028$ \\ 
$5929.049$ &$18.571 \pm  0.027$ &$18.162 \pm  0.021$ \\ 
$5931.046$ &$18.520 \pm  0.026$ &$18.203 \pm  0.022$ \\ 
$5932.046$ &$18.560 \pm  0.025$ &$18.193 \pm  0.021$ \\ 
$5933.046$ &$18.557 \pm  0.025$ &$18.194 \pm  0.021$ \\ 
$5945.038$ &$18.524 \pm  0.027$ &$18.119 \pm  0.022$ \\ 
$5946.030$ &$18.546 \pm  0.026$ &$18.176 \pm  0.022$ \\ 
$5955.988$ &$18.541 \pm  0.027$ &$18.177 \pm  0.023$ \\ 
$5957.052$ &$18.509 \pm  0.029$ &$18.164 \pm  0.024$ \\ 
$5959.050$ &$18.526 \pm  0.028$ &$18.130 \pm  0.023$ \\ 
$5960.050$ &$18.498 \pm  0.030$ &$18.109 \pm  0.025$ \\ 
$5963.050$ &$18.455 \pm  0.023$ &$18.093 \pm  0.020$ \\ 
$5971.008$ &$18.483 \pm  0.035$ &$18.078 \pm  0.028$ \\ 
$5979.929$ &$18.481 \pm  0.026$ &$18.096 \pm  0.021$ \\ 
$5983.010$ &$18.524 \pm  0.029$ &$18.083 \pm  0.022$ \\ 
$5986.995$ &$18.507 \pm  0.028$ &$18.059 \pm  0.022$ \\ 
$5987.671$ &$18.495 \pm  0.035$ &$18.058 \pm  0.028$ \\ 
$5999.900$ &$18.572 \pm  0.033$ &$18.079 \pm  0.024$ \\ 
$6000.801$ &$18.532 \pm  0.023$ &$18.101 \pm  0.018$ \\ 
$6007.905$ &$18.577 \pm  0.030$ &$18.138 \pm  0.025$ \\ 
$6008.890$ &$18.564 \pm  0.026$ &$18.088 \pm  0.020$ \\ 
$6011.836$ &$18.578 \pm  0.026$ &$18.086 \pm  0.019$ \\ 
$6012.958$ &$18.502 \pm  0.034$ &$18.108 \pm  0.028$ \\ 
$6013.904$ &$18.527 \pm  0.025$ &$18.125 \pm  0.020$ \\ 
$6014.855$ &$18.517 \pm  0.023$ &$18.116 \pm  0.019$ \\ 
$6015.829$ &$18.511 \pm  0.023$ &$18.120 \pm  0.019$ \\ 
$6017.719$ &$18.533 \pm  0.031$ &$18.120 \pm  0.024$ \\ 
$6029.686$ &$18.496 \pm  0.023$ &$18.127 \pm  0.019$ \\ 
$6030.756$ &$18.576 \pm  0.025$ &$18.099 \pm  0.019$ \\ 
$6035.876$ &$18.551 \pm  0.030$ &$18.159 \pm  0.024$ \\ 
$6036.763$ &$18.567 \pm  0.028$ &$18.200 \pm  0.023$ \\ 
$6038.694$ &$18.566 \pm  0.025$ &$18.136 \pm  0.019$ \\ 
$6039.849$ &$18.573 \pm  0.030$ &$18.108 \pm  0.023$ \\ 
$6041.695$ &$18.567 \pm  0.025$ &$18.153 \pm  0.020$ \\ 
$6049.845$ &$18.470 \pm  0.063$ &$18.173 \pm  0.057$ \\ 
$6051.684$ &$18.536 \pm  0.047$ &$18.093 \pm  0.036$ \\ 
$6052.683$ &$18.578 \pm  0.052$ &$18.105 \pm  0.039$ \\ 
$6053.684$ &$18.665 \pm  0.064$ &$18.128 \pm  0.046$ \\ 
$6054.688$ &$18.528 \pm  0.041$ &$18.072 \pm  0.032$ \\ 
$6055.717$ &$18.612 \pm  0.039$ &$18.109 \pm  0.029$ \\ 
$6057.760$ &$18.642 \pm  0.038$ &$18.092 \pm  0.028$ \\ 
$6059.686$ &$18.534 \pm  0.025$ &$18.110 \pm  0.020$ \\ 
$6060.668$ &$18.549 \pm  0.024$ &$18.099 \pm  0.018$ \\ 
$6066.656$ &$18.574 \pm  0.026$ &$18.142 \pm  0.020$ \\ 
$6070.655$ &$18.562 \pm  0.025$ &$18.087 \pm  0.019$ \\ 
$6071.655$ &$18.587 \pm  0.028$ &$18.106 \pm  0.021$ \\ 
$6072.655$ &$18.623 \pm  0.032$ &$18.059 \pm  0.022$ \\ 
$6077.661$ &$18.636 \pm  0.041$ &$18.085 \pm  0.029$ \\ 
$6082.659$ &$18.575 \pm  0.059$ &$18.121 \pm  0.045$ \\ 

\enddata
\tablecomments{The first point is the original SDSS observation from 2004 December \citep{Inada2006}. 
    The Keplercam observations started in 2007 January. The 
   uncertainties are dominated by the noise in the individual epochs rather than the photometric
   calibration.
   }
\label{tab:lc}
\end{deluxetable}

\end{document}